\documentclass[prd,aps,showpacs,showkeys]{revtex4}

\usepackage{amssymb}
\usepackage{amsmath}
\usepackage{latexsym}

\usepackage[dvips]{graphicx}

\newcommand{\sikib}{\begin{eqnarray}}
\newcommand{\sikie}{\end{eqnarray}}
\newcommand{\sikibnon}{\begin{eqnarray*}}
\newcommand{\sikienon}{\end{eqnarray*}}

\newcommand{\gin}[1]{g_{in}(\vec{#1})}
\newcommand{\gout}[1]{g_{out}(\vec{#1})}

\newcommand{\angl}{\theta_p}
\newcommand{\anglm}{\theta_m}
\newcommand{\rt}{\tilde{t}}
\newcommand{\rtau}{\tilde{\tau}}

\begin{document}

\title{The generalised second law and the black hole evaporation in an empty space as a nonequilibrium process}
\author{Hiromi Saida}
\email{saida@daido-it.ac.jp}
\affiliation{Department of Physics, Daido Institute of Technology, \\
Takiharu-cho 10-3, Minami-ku, Nagoya 457-8530, Japan}


\begin{abstract}
When a black hole is in an empty space on which there is no matter field except that of the Hawking radiation (Hawking field), then the black hole evaporates and the entropy of the black hole decreases. The generalised second law guarantees the increase of the total entropy of the whole system which consists of the black hole and the Hawking field. That is, the increase of the entropy of the Hawking field is faster than the decrease of the black hole entropy. In naive sense, one may expect that the entropy increase of the Hawking field is due to the self-interaction among the composite particles of the Hawking field, and that the {\it self}-relaxation of the Hawking field results in the entropy increase. Then, when one consider a non-self-interacting matter field as the Hawking field, it is obvious that the self-relaxation does not take place, and one may think that the total entropy does not increase. However, using nonequilibrium thermodynamics which has been developed recently, we find for the non-self-interacting Hawking field that the rate of entropy increase of the Hawking field (the entropy emission rate by the black hole) grows faster than the rate of entropy decrease of the black hole along the black hole evaporation in the empty space. The origin of the entropy increase of the Hawking field is the increase of the black hole temperature. Hence an understanding of the generalised second law in the context of the nonequilibrium thermodynamics is suggested; even if the self-relaxation of the Hawking field does not take place, the temperature increase of the black hole during the evaporation process causes the entropy increase of the Hawking field to result in the increase of the total entropy.
\end{abstract}



\maketitle

\section{Introduction}
\label{sec-intro}

When a black hole is in an empty space on which there is no matter field except that of the Hawking radiation (Hawking field) \cite{ref-hr}, then an outgoing energy flow from the black hole into the empty space arises due to the Hawking radiation and the black hole loses its mass energy. This denotes that a black hole evaporates. Concerning the black hole evaporation in the empty space, the generalised second law (GSL) of the black hole thermodynamics is an important and interesting issue \cite{ref-gsl} \cite{ref-bht}. The GSL conjectures that the entropy of a black hole $S_g$ is given by the horizon area (divided by 4) and that the total entropy of the whole system, which consists of the black hole and the Hawking field, increases during the black hole evaporation process. Note that, when the GSL holds, the black hole entropy $S_g$ decreases during the evaporation process, but the entropy of the Hawking field increases faster than the decrease of $S_g$. 

By the way, the outgoing energy flow from the black hole into the empty space denotes that the Hawking field is in a nonequilibrium state, since no energy flow arises in an equilibrium state. Therefore the black hole evaporation in the empty space should be interpreted as a nonequilibrium process of the whole system. In order to prove the GSL in the context of the black hole evaporation in the empty space, we have to define a nonequilibrium entropy of the Hawking field. 

However, because a general definition of the nonequilibrium entropy of matters has not yet been formulated, the existing proofs of the GSL consider an equilibrium between a black hole and a heat bath surrounding it, or assume the existence of a well-defined nonequilibrium entropy of an arbitrary matter \cite{ref-gsl.2}. Here note that, in the existing proofs, a self-interacting matter field has been considered as the Hawking field. For the black hole evaporation in the empty space, the self-interaction causes the {\it self}-relaxation of the Hawking field itself during propagating on the empty space to increase the entropy of the Hawking field (if a nonequilibrium entropy were to be defined). Then one may expect naively that the self-relaxation of the Hawking field is the origin of the increase of the total entropy of the whole system. This seems to be a very reasonable picture of the GSL in the context of the black hole evaporation in the empty space. However, I dare to suggest an understanding of the GSL in the context of the nonequilibrium thermodynamics as follows. {\it The self-interaction of the Hawking field is not necessary, but the increase of the black hole temperature during the evaporation process plays an essential role for the validity of the GSL. That is, due to the temperature increase, the rate of entropy increase of the Hawking field (the entropy emission rate by the black hole) grows faster than the rate of entropy decrease of the black hole, then the total entropy increases without the self-interaction of the Hawking field.} Note that the increase of the black hole temperature along the loss of mass energy is due to the negative heat capacity of the black hole. The negative heat capacity is peculiar to self-gravitating systems \cite{ref-sgs}. Therefore our understanding of the GSL depends strongly on the feature of the self-gravitation of the black hole.

In order to explain our understanding of the GSL, it is appropriate to consider a non-self-interacting matter field as the Hawking field. For example, photon, graviton, and a massless free Klein-Gordon field ($\square \Phi = 0$) can be the non-self-interacting Hawking fields. Recently, in the study on nonequilibrium thermodynamics of a non-self-interacting matter field, it has been reported that a nonequilibrium thermodynamics for a radiation field (photon gas) was formulated in a consistent way \cite{ref-sst}. With referring to report \cite{ref-sst}, this paper aims to calculate the time evolution of the total entropy of a black hole and the non-self-interacting Hawking field in a simplified model of the black hole evaporation in the empty space, and to give an understanding of the GSL as explained above. Further, since report \cite{ref-sst} treats the photons which are collisionless and massless particles, we consider that the Hawking field is not only non-self-interacting but also massless.

This paper is organised as follows. The strategy to reach our understanding of the GSL is explained in section \ref{sec-strategy}. In section \ref{sec-sst}, the nonequilibrium entropy obtained in \cite{ref-sst} is briefly explained. Section \ref{sec-entropy} is devoted to the application of the nonequilibrium entropy to our strategy. Summary and discussions are in section \ref{sec-sd}.

We set the fundamental constants unity, $\hbar = G = c = k_B = 1$. Then the Stefan-Boltzmann constant is $\sigma = \pi^2/60$, which is appropriate for a photon gas. When one consider a general non-self-interacting massless field as the Hawking field, replace the Stefan-Boltzmann constant as $\sigma \to \sigma^{\prime} = N\, \pi^2/120 = N\, \sigma/2$, where $N = n_b + ( 7/8 )\, n_f$, and $n_b$ ($n_b = 2$ for photons) is the number of helicities of massless bosonic field and $n_f$ is that of massless fermionic field.

\section{Strategy to reach our understanding of the GSL}
\label{sec-strategy}

\subsection{GSL in the context of the black hole evaporation}
\label{sec-strategy.gsl}

Before proceeding to the explanation of our strategy to reach the understanding of the GSL in the context of nonequilibrium thermodynamics, we summarise the statement of the GSL in the context of the black hole evaporation in the empty space.

Some classical properties of a black hole solution of the Einstein equation can be mathematically formulated so that they correspond to the four laws of the ordinary thermodynamics \cite{ref-bht}. On the other hand, the Hawking radiation, which is a quantum effect of matter fields on a classical black hole spacetime, implies that the black hole emits a thermal radiation of temperature $T_g = \kappa/2\pi$, where $\kappa$ is the surface gravity at the black hole horizon \cite{ref-hr}. Then, because of the Hawking radiation, the mathematical correspondence between the properties of a black hole and the laws of thermodynamics seems to be physically reasonable. This mathematical correspondence has been known as the black hole thermodynamics. In the black hole thermodynamics, the quantity $S_g$ which corresponds to an entropy of the black hole is determined to be $S_g = A_g/4$, where $A_g$ is the spatial area of the black hole horizon. 

Hereafter we consider the Schwarzschild black hole for simplicity. Consider the case that the black hole is in an empty space on which no matter field except the Hawking field exists, and the black hole evaporates due to the Hawking radiation. The black hole thermodynamics suggests that the black hole behaves like a spherical black body whose equations of states are 
\sikib
 E_g = \frac{1}{8 \, \pi T_g} = \frac{R_g}{2} \quad , \quad
 S_g = \frac{1}{16 \, \pi T_g^{\,2}} = \pi R_g^{\,2} \, ,
\label{eq-strategy.eos}
\sikie
where $R_g$ and $E_g$ are respectively the areal radius and the mass energy of the black hole. 

When the Hawking radiation (the quantum effect of matter fields) is not considered, the mass energy $E_g$ never decrease and the quantity $S_g$ never decrease as well. Then, from classical viewpoint, it seems that $S_g$ corresponds to the entropy of the black hole. However it is obvious that, when the Hawking radiation is taken into account, $S_g$ decreases during the black hole evaporation process due to the energy loss by the Hawking radiation. Then the GSL conjectures that the quantity $S_g$ in equations \eqref{eq-strategy.eos} is the true entropy of the black hole, and the total entropy of the whole system increases during the black hole evaporation process in the empty space \cite{ref-gsl}.

Here we give some supplementary comments on the energetic and entropic properties of the black hole. For the first on the energetics, the equations of states \eqref{eq-strategy.eos} give the negative heat capacity $C_g$ of the black hole,
\sikib
 C_g = \frac{dE_g}{dT_g} = - \frac{1}{8 \, \pi T_g^{\,2}} = - 2 \, \pi R_g^{\,2} < 0 \, .
\label{eq-strategy.capa}
\sikie
This denotes that, when the energy $E_g$ is lost, the temperature $T_g$ increases. The negative heat capacity is a peculiar energetic property to the self-gravitating systems \cite{ref-sgs}. Therefore it is obvious that the equations of states \eqref{eq-strategy.eos} contain the self-gravitational effects of the black hole on its own thermodynamic states. Next for the black hole entropy, it has already been revealed that, if there is no black hole horizon ($R_g = 0$) on the spacetime, the quantity which corresponds to the entropy of the spacetime itself is zero \cite{ref-gh} \cite{ref-add}. That is, the quantity which corresponds to the entropy of the whole gravitational field is given by the quantity $S_g$ in equations \eqref{eq-strategy.eos}. Hence, the energetic and entropic properties of a black hole are well encoded in the equations of states \eqref{eq-strategy.eos}.

\subsection{Strategy to reach our understanding of the GSL}
\label{sec-strategy.strategy}

We have to define the total entropy of the whole system. It has already been shown in reference \cite{ref-add} that, for the equilibrium state of the black hole with a general self-interacting Hawking field, the total entropy is given by a simple sum of the black hole entropy $S_g$ with the equilibrium entropy of the general Hawking field. Therefore, in the case of the black hole evaporation in the empty space, we assume that the equilibrium case is extended simply so that the total entropy $S_{tot}$ of the whole system which consists of the black hole and the nonequilibrium Hawking field is given by the simple sum, 
\sikib
 S_{tot} = S_g + S_m \, ,
\label{eq-strategy.total.general}
\sikie
where $S_g$ is given by equations \eqref{eq-strategy.eos} and $S_m$ is the nonequilibrium entropy of the general self-interacting Hawking field. The GSL conjectures the inequality $dS_{tot} > 0$ during the black hole evaporation process. Here note that, because the black hole evaporation proceeds very slowly (as seen later in section \ref{sec-entropy.model}), we approximate the black hole itself is in an equilibrium state at each moment during the evaporation process, while the Hawking field is in a nonequilibrium state. Hence we use the {\it equilibrium} quantity $S_g$ as the black hole entropy in equation \eqref{eq-strategy.total.general}. More discussions on the validity of the additivity \eqref{eq-strategy.total.general} and of the existence of a well-defined nonequilibrium entropy $S_m$ of an arbitrary matter are given later in section \ref{sec-sd}. 

Because the Hawking field is in a nonequilibrium state due to the energy flow by the Hawking radiation, the problem is how to deal with the nonequilibrium entropy $S_m$. Note that, it is appropriate to assume that the nonequilibrium entropy of any self-interacting matter increases during a relaxation process, since it is the self-interaction that causes the increase of the matter entropy in the context of the ordinary equilibrium thermodynamics. Therefore, if we find the increase of the total entropy of the whole system which consists of the black hole and the {\it non}-self-interacting Hawking field, then it implies that the GSL holds with an arbitrary self-interacting Hawking field. Here one may show an example as an objection that, when an inter-stellar gas collapses to form a star, although the net entropy will increase, the self-gravitational effect decreases the entropy of the collapsing gas. However as discussed in appendix \ref{app-star}, we can expect that this objection is not true of the Hawking field. 

Next turn our attention from self-interaction to the effect by the {\it external}-gravitational field on the Hawking radiation. It should be noted that, even when the self-interaction (including self-gravitation) of the Hawking field is ignored, there remain some gravitational interactions between the Hawking field and the black hole, which are for example the gravitational red shift, curvature scattering, lens effect and so on. The external-gravitational field works as if a virtual medium on which the Hawking field propagates, then the composite particles of the Hawking field interact with the virtual medium to result in the relaxation of the Hawking field toward some equilibrium state. Therefore, it is appropriate to assume that the external-gravitational interaction causes the increase of the entropy of the Hawking field. Hence we expect that, if we can calculate a nonequilibrium entropy of the non-self-interacting Hawking field by ignoring the external-gravitational interaction, then the rate of increase of such an approximate entropy is smaller than that of the nonequilibrium entropy of the non-self-interacting Hawking field calculated by including the external-gravitational interaction. 

From the above discussions, we can expect that the following inequality holds along the black hole evaporation process in the empty space,
\sikib
 dS_{tot} > dS_{tot}^{(min)} \, ,
\label{eq-strategy.ineq}
\sikie
where $S_{tot}$ is the total entropy with a general self-interacting Hawking field given by equation \eqref{eq-strategy.total.general} which includes the gravitational interactions between the black hole and the Hawking field, and $S_{tot}^{(min)}$ is the total entropy of the whole system which consists of the black hole and the non-self-interacting Hawking field. Here the entropy $S_{tot}^{(min)}$ is decomposed as
\sikib
 S_{tot}^{(min)} = S_g + S_{nsi} \, ,
\label{eq-strategy.total.rad}
\sikie
where $S_{nsi}$ is the nonequilibrium entropy of the non-self-interacting Hawking field without including the gravitational interaction with the black hole. Therefore, if the inequality $dS_{tot}^{(min)} > 0$ holds, then we can expect that the GSL, $dS_{tot} > 0$, follows. 

In naive sense, one may think that the nonequilibrium entropy $S_{nsi}$ of the non-self-interacting Hawking field does not increase, because the self-interactions and the external-gravitational potential are ignored and the Hawking field never relaxes by itself. The absence of the self-relaxation is a right conclusion from above discussions. However it should be emphasised that there still remains an interaction, the emission process of the Hawking radiation by the black hole. Here note that the negative heat capacity $C_g$ of the black hole denotes the increase of the black hole temperature $T_g$ along the loss of the mass energy $E_g$. Therefore it can be expected that the emission process causes the increase of the entropy $S_{nsi}$ of the non-self-interacting Hawking field even when the gravitational interaction with the black hole is ignored. The higher the temperature $T_g$ grows, the more amount of matter entropy is emitted with the Hawking radiation.

In this paper, by constructing a concrete formula of the nonequilibrium entropy $S_{nsi}$ of a non-self-interacting Hawking field without including the gravitational interaction with the black hole, we aim to show that the inequality $dS_{tot}^{(min)} > 0$ holds during the black hole evaporation in the empty space. The concrete method for the calculation of the entropy $S_{nsi}$ is explained in the following sections.

\subsection{The model of the black hole evaporation}
\label{sec-strategy.sst}

Because we are interested in the non-self-interacting Hawking field without the gravitational interaction with the black hole, it is enough to consider the model that a {\it black body} satisfying the equations of states \eqref{eq-strategy.eos} is put in a flat spacetime. This modelling retains the self-gravitational effects of the black hole on its own thermodynamic states (negative heat capacity), as mentioned in section \ref{sec-strategy.gsl}. Therefore, in order to pick up the nonequilibrium effects of the Hawking field on the black hole evaporation process, we consider a simplified model named NHR after the nonequilibrium Hawking radiation: 
\begin{description}
\item[NHR Model:] Consider a spherical black body satisfying equations \eqref{eq-strategy.eos}, which emits some non-self-interacting massless matter fields with Planckian spectrum (free Hawking fields). Then put the black body in a flat spacetime on which there exists no matter field except the free Hawking fields. Then the black body evaporates due to equations \eqref{eq-strategy.eos} and the emission of the free Hawking fields. This evaporation can be understood as a relaxation process of the whole system which consists of the black body and the fee Hawking fields.
\end{description}
Here note that, although the self-interactions (including the self-gravitation) of the Hawking field are ignored, however the self-gravitational effects of the black hole on its own thermodynamic state is encoded in the equations of states \eqref{eq-strategy.eos}. Henceforth we call the black body {\it the black hole}, and call the non-self-interacting massless fields {\it the free Hawking fields}. 

In the NHR model, the black hole continues to emit the free Hawking fields without any in-falling energy into the black hole, and the free Hawking fields continue to spread out into the empty space. That is, an outgoing energy flow from the black hole into the empty space exists during the black hole evaporation process. The outgoing energy flow equals the rate of energy increase in the free Hawking fields.

Further because of the quasi-equilibrium approximation of the black hole itself (see the first paragraph in section \ref{sec-strategy.strategy}), we notice that, from the thermodynamic viewpoint, the thermodynamic state of the black hole itself during the evaporation process changes along a sequence of equilibrium states characterised by equations \eqref{eq-strategy.eos} in the state space, and that the black hole behaves always as a perfect black body of temperature $T_g$ at each moment during the evaporation process. (Recall the quasi-static process in the ordinary equilibrium thermodynamics.) Therefore the rate of energy increase $J$ in the free Hawking fields in the NHR model is given by the Stefan-Boltzmann law,
\sikib
 J = \sigma^{\prime} \, T_g^{\,4} \, A_g \, ,
\label{eq-strategy.flux}
\sikie
where $\sigma^{\prime}$ is the general Stefan-Boltzmann constant given at the end of section \ref{sec-intro}.

This $J$ is equivalent to the energy flow in the free Hawking fields, and denotes that, while the thermodynamic state of the black hole changes along a sequence of equilibrium states, the thermodynamic state of the free Hawking fields should change along a sequence of nonequilibrium states in the state space, where each nonequilibrium state in the sequence of the field's states should have the energy flow $J$. Hence the time evolution of the total entropy $S_{tot}^{(min)}$ should be analysed with a nonequilibrium thermodynamics for the free Hawking fields. In order to carry out the nonequilibrium analysis, we refer report \cite{ref-sst} which has formulated a nonequilibrium thermodynamics for a radiation field (photon gas) in a consistent way. This can easily be extended to a general non-self-interacting massless field as shown in the next section \ref{sec-sst}. Then the nonequilibrium entropy $S_{nsi}$ of the free Hawking fields is concretely given using the formula obtained in \cite{ref-sst}. 

Here one may point out that, for an equilibrium of a black hole with the Hawking field, the gravitational red shift results in the need for a higher temperature of the Hawking field than the black hole temperature $T_g$ in order to retain the equilibrium against the attractive gravitational force by the black hole \cite{ref-rd}. Then an energy flow stronger than $J$ would be appropriate in order to retain the effect of the gravitational red shift. However the NHR model ignores the external-gravitational potential, and the energy flow $J$ of equation \eqref{eq-strategy.flux} is suitable for our purpose. Further note that a lower temperature of the free Hawking fields implies a smaller emission rate of entropy by the black hole. Therefore, by ignoring the gravitational red shift, we can recognise that inequality \eqref{eq-strategy.ineq} becomes more reasonable.

Although some simplifications and assumptions are put in the NHR model, which one may think as a too simplified model, we will use it to obtain some insight into the GSL. Further in order to avoid confusions, it is helpful for readers to mention here that the so-called black hole phase transition (or the Hawking-Page phase transition for AdS black holes) in the ordinary black hole thermodynamics \cite{ref-hp} is out of the scope of this paper. The black hole phase transition is the conclusion which comes from the study on the equilibrium states of the system in which a black hole is put in a heat bath. Here in this paper we do not consider any heat bath surrounding the black hole, but put a black hole in an empty and infinitely large space where the black hole can not be in any equilibrium with its environment. We are interested in the dynamical process of a black hole evaporation in the empty space as a nonequilibrium process, and aim to clarify the effect of the increase of black hole temperature on the GSL.

\section{Steady state thermodynamics for free Hawking fields}
\label{sec-sst}

\begin{figure}[t]
 \begin{center}
 \includegraphics[height=25mm]{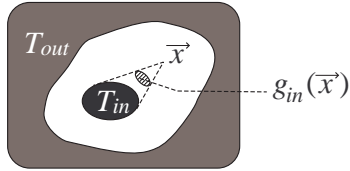}
 \end{center}
\caption{MS model. The radiation field between the outer and inner black bodies is in a steady state. The entropy of the radiation field is given by the steady state thermodynamics for a radiation field.}
\label{pic-1}
\end{figure}

As a middle step before proceeding to the calculation of the nonequilibrium entropy $S_{nsi}$ of the free Hawking fields in the NHR model, we consider a more simplified model which considers a radiation field (photon gas) as the free Hawking fields and can be realised in a laboratory experiment. This simple model is shown in figure \ref{pic-1}, and named MS after the middle step:
\begin{description}
\item[MS model:] Make a vacuum cavity in a large black body of equilibrium temperature $T_{out}$ and put an another smaller black body of temperature $T_{in} \, ( \neq T_{out} )$ in the cavity. For the case $T_{in} > T_{out}$, the radiation field (photon gas) emitted by the two black bodies causes an energy flow from the inner black body to the outer one. When the outer black body is isolated from the outside world, the whole system which consists of the two black bodies and the radiation field between them relaxes to a total equilibrium state in which the two black bodies and the radiation field have the same equilibrium temperature.
\end{description}
Note that the equations of states of the two black bodies are not specified. Here following report \cite{ref-sst}, we consider that the black bodies emit a radiation field (photon gas). However it is easily extended to a general non-self-interacting massless field by replacing the Stefan-Boltzmann constant $\sigma$ with the generalized one $\sigma^{\prime}$ as mentioned at the end of section \ref{sec-intro}. In section \ref{sec-entropy}, we will extend the MS model to the NHR model. To do so, we will make the inner black body have the equations of states \eqref{eq-strategy.eos} and remove the outer black body by setting the cavity infinitely large and $T_{out} = 0$.

Concerning the relaxation process required in the MS model, consider the case satisfying the following two conditions. The first is that the thermodynamic state of each black body evolves so slowly that it changes along a sequence of equilibrium states in the state space during the relaxation process. The second condition is that the volume of the cavity is so small that the speed of light is approximated as infinity during the relaxation process. These two conditions denote that \cite{ref-sst}, at each moment of the relaxation process, the retarded effect of photons is ignored, and the thermodynamic state of the radiation field sandwiched by the two black bodies is well approximated by the nonequilibrium state which possesses a {\it stationary} energy flow $J_{rad}$ determined by the temperatures of the two black bodies $T_{in}$ and $T_{out}$. The energy flow $J_{rad}$ is given by the Stefan-Boltzmann law, $J_{rad} = \sigma \, (\, T_{in}^{\,4} - T_{out}^{\,4} \, ) \, A_{in}$, where $A_{in}$ is the surface area of the inner black body. This $J_{rad}$ equals the energy exchange par a unit time between the two black bodies via the radiation field. 

From above discussions, it is enough for us to consider a set (state space) of the {\it steady states} (macroscopically stationary nonequilibrium state) of the radiation field whose nonequilibrium nature is raised by the steady energy flow $J_{rad}$. That is, the sequence of the nonequilibrium states of the radiation field during the relaxation process lies in the set (state space) of the steady states. Although each steady state alone does not cause any time evolution of the system, however the time evolution of the temperatures $T_{in}$ and $T_{out}$ can be obtained by shifting the thermodynamic state of the radiation field along the sequence of the steady states as time goes on. Therefore we need a consistent definition of the steady state entropy of the radiation field. Once the steady state entropy of the radiation field is obtained, we can express explicitly the total entropy during the relaxation process of the whole system which consists of the two black bodies and the radiation field. Further, by extending the definition of the steady state entropy into the case of an infinitely large cavity with $T_{out} = 0$, it is expected that the total entropy $S_{tot}^{(min)}$ of the NHR model can be analysed explicitly as well. 

A consistent thermodynamic framework for the steady states for a radiation field in the MS model has already been constructed \cite{ref-sst}. The outline of the construction of the entropy in the steady state thermodynamics for a radiation field in the MS model is as follows. 

Refer the Landau-Lifshitz type definition of a nonequilibrium entropy for a bosonic gas \cite{ref-ll},
\sikib
 S_{boson} =
  \int \frac{dp^3}{(2\, \pi)^3} dx^3 \, g_{\vec{p},\vec{x}}
  \left[\, \left(\, 1 + f_{\vec{p},\vec{x}} \,\right) \,
             \ln\left(\, 1 + f_{\vec{p},\vec{x}} \,\right)
         - f_{\vec{p},\vec{x}} \, \ln f_{\vec{p},\vec{x}}
  \,\right] \, ,
\label{eq-sst.ll.boson}
\sikie
where $\vec{p}$ is the momentum of a photon, $\vec{x}$ is a spatial point, and $g_{\vec{p},\vec{x}}$ and $f_{\vec{p},\vec{x}}$ are respectively the number of states and the average number of particles at a point $(\vec{p},\vec{x})$ in the phase space of the bosonic gas. Note that it has also been shown in reference \cite{ref-ll} that the maximisation of $S_{boson}$ for an isolated system ($\delta S_{boson} = 0$) gives the equilibrium Bose distribution. This is frequently referred to in many works on nonequilibrium systems as the {\it H-theorem}. However in reference \cite{ref-ll}, concrete forms of $g_{\vec{p},\vec{x}}$ and $f_{\vec{p},\vec{x}}$ are not specified, since an arbitrary system is considered.

In the MS model, the radiation field is sandwiched between two black bodies. Then we can determine $g_{\vec{p},\vec{x}}$ and $f_{\vec{p},\vec{x}}$ as
\sikib
 g_{\vec{p},\vec{x}} = 2 \,\,\text{(helicities of a radiation field)} \quad , \quad
 f_{\vec{p},\vec{x}} = \frac{1}{\exp[\omega / T(\vec{p},\vec{x})] - 1} \, ,
\label{eq-sst.distribution}
\sikie
where the frequency of a photon $\omega = \left| \vec{p} \right|$, and $T(\vec{p},\vec{x})$ is given by
\sikibnon
 T(\vec{p},\vec{x}) =
 \begin{cases}
  T_{in}  & \text{for} \,\,\, \vec{p} = \vec{p}_{in} \,\,\, \text{at $\vec{x}$} \\
  T_{out} & \text{for} \,\,\, \vec{p} = \vec{p}_{out} \,\,\, \text{at $\vec{x}$}
 \end{cases} \, ,
\sikienon
where $\vec{p}_{in}$ is the momentum of the photon emitted by the inner black body and $\vec{p}_{out}$ by the outer black body. The $\vec{x}$-dependence of $T(\vec{p},\vec{x})$ arises, because the directions in which the photons of $\vec{p}_{in}$ and $\vec{p}_{out}$ can come to a point $\vec{x}$ vary from point to point. 

From above, we obtain the steady state entropy of a radiation field, $S_{rad}$,
\sikib
 S_{rad} = \int dx^3 s_{rad}(\vec{x}) \quad , \quad
 s_{rad}(\vec{x}) = \frac{16 \, \sigma}{3} \,
  \left(\, \gin{x}\, T_{in}^{\,3} + \gout{x}\, T_{out}^{\,3} \,\right) \, ,
\label{eq-sst.entropy.rad}
\sikie
where $\gin{x}$ is the solid angle (divided by $4 \pi$) covered by the direction of $\vec{p}_{in}$ at $\vec{x}$ as shown in figure \ref{pic-1}, and $\gout{x}$ is similarly defined with $\vec{p}_{out}$. In deriving \eqref{eq-sst.entropy.rad}, we used the relations, $\int^{\infty}_0 dx \, \left[ x^2/( e^x - 1 )\right] \, \ln( e^x - 1 ) = 11 \pi^4/ 180$ and $\int^{\infty}_0 dx \, \left[ x^2/( 1 - e^{-x} )\right] \, \ln( 1 - e^{-x} ) = -\pi^4/ 36$ .

Further, with defining the steady state internal energy $E_{rad} = 2 \int dp^3 dx^3 \, \omega \, f_{\vec{p},\vec{x}}$ and the other variables such as the free energy with somewhat careful discussions, it has already been checked that the 0th, 1st, 2nd and 3rd laws of the ordinary equilibrium thermodynamics are extended to include the steady states of a radiation field. That is, the steady state thermodynamics for a radiation field has already been constructed in a consistent way. Especially in the steady state entropy \eqref{eq-sst.entropy.rad}, it has already been revealed in \cite{ref-sst} that the total entropy of the whole system which consists of the two black bodies and the radiation filed increases monotonically during the relaxation process required in the MS model, $dS_{in} + dS_{out} + dS_{rad} \ge 0$, where the equality holds for the total equilibrium state, $S_{in}$ and $S_{out}$ are the entropies of the two black bodies which are given by the ordinary equilibrium thermodynamics, and it is assumed that the two black bodies are made of ordinary materials and have positive heat capacities.

When we consider a massless fermionic field instead of the radiation field in the MS model, equation \eqref{eq-sst.ll.boson} is replaced by \cite{ref-ll}
\sikib
 S_{fermion} =
 -\int \frac{dp^3}{(2\, \pi)^3} dx^3 \, g_{\vec{p},\vec{x}}
  \left[\, \left(\, 1 - f_{\vec{p},\vec{x}} \,\right) \,
             \ln\left(\, 1 - f_{\vec{p},\vec{x}} \,\right)
         + f_{\vec{p},\vec{x}} \, \ln f_{\vec{p},\vec{x}}
  \,\right] \, ,
\label{eq-sst.ll.fermion}
\sikie
where $g_{\vec{p},\vec{x}} = n_f$ is the effective number of helicities, and $f_{\vec{p},\vec{x}} = [\,\exp[\omega / T(\vec{p},\vec{x})] + 1\,]^{-1}$. Then, following the same procedure as a radiation field, we can obtain the steady state entropy of a massless fermionic field as
\sikib
 S_f = \frac{7}{8} \, \frac{n_f}{2}\, S_{rad} \, ,
\label{eq-sst.entropy.fermion}
\sikie
where we used, $\int^{\infty}_0 dx \, \left[ x^2/( e^x + 1 )\right] \, \ln( e^x + 1 ) = \pi^4/ 60 + A$, and $\int^{\infty}_0 dx \, \left[ x^2/( 1 + e^{-x} )\right] \, \ln( 1 + e^{-x} ) = 11 \pi^4/180 - A$, where $A = [ (\ln 2)^2 - \pi^2 ]\,(\ln 2)^2/6 + 4 \, \phi(4,1/2) + (7 \ln 2/2)\, \zeta(3)$, $\zeta(z)$ is the zeta function and $\phi(z,s)$ is the modified zeta function (Apell's function). Hence, when the black bodies in the MS model emit $n_f$ massless fermionic modes and $n_b$ massless bosonic modes ($n_b = 2$ for a radiation field), the steady state entropy of the matter fields between the two black bodies is given by equation \eqref{eq-sst.entropy.rad} by replacing $\sigma$ by $\sigma^{\prime}$ given at the end of section \ref{sec-intro}.

\section{Time evolution of the total entropy}
\label{sec-entropy}

\subsection{From the MS model to the NHR model}
\label{sec-entropy.model}

We proceed to the analysis of the total entropy $S_{tot}^{(min)}$ in the NHR model, which can be obtained from the MS model by considering the equations of states of the inner black body as equations \eqref{eq-strategy.eos} and by removing the outer black body. The way of removing the outer black body from the MS model is to set the cavity's volume infinite and to set the outer temperature zero $T_{out} = 0$. In the NHR model, the free Hawking fields continue to spread out into the empty space.

The time evolution of the black hole radius $R_g$ is given by equations \eqref{eq-strategy.eos} and \eqref{eq-strategy.flux},
\sikib
 \frac{d E_g}{dt} = - J \quad \Rightarrow \quad
 R_g(t) = R_0 \, \left( 1 - \frac{N\,t}{1280 \, \pi \, R_0^3} \right)^{1/3}
        = \frac{1}{4 \, \pi \, T_g(t)} \, ,
\label{eq-entropy.radius}
\sikie
where $R_0 = R_g(0)$ is the initial radius, and it is assumed that the emission of the Hawking radiation starts at $t=0$. This $R_g(t)$ leads the time evolution of the thermodynamic state of the black hole and the steady state of the free Hawking fields. Equation \eqref{eq-entropy.radius} gives the evaporation time (life time) of the black hole,
\sikib
 t_{ev} = 1280\,\pi\,\frac{R_0^3}{N} \, .
\label{eq-entropy.life.time}
\sikie
On the other hand, the time scale $t_{bh}$ that a massless particle travels across the size of the black hole is given by $t_{bh} = R_0$. This time $t_{bh}$ gives a typical time scale of the free Hawking fields to spread out into the empty space. Therefore, when the inequality $t_{ev}/t_{bh} \gg O(1)$ holds, the black hole evaporation proceeds very slowly in comparison with the time scale $t_{bh}$. Henceforth, we assume the order relation $t_{ev}/t_{bh} \gg O(1)$, which denotes that the quasi-equilibrium approximation of the black hole itself is valid as mentioned in section \ref{sec-strategy.strategy}. More discussion on this assumption is given later in section \ref{sec-sd}.

\subsection{Nonequilibrium entropy of the free Hawking fields}
\label{sec-entropy.hr}

In order to apply the steady state thermodynamics \cite{ref-sst} to the free Hawking fields in the NHR model, the distribution function of photons \eqref{eq-sst.distribution} should be modified to include the retarded effect of composite particles ($n_b$ bosons and $n_f$ fermions) travelling in an infinitely large space. In this section, we construct the modified distribution function for the first, then obtain the nonequilibrium entropy of the free Hawking fields in the NHR model by substituting the modified distribution function into formulae \eqref{eq-sst.ll.boson} and \eqref{eq-sst.ll.fermion}. Here note that, because of the quasi-equilibrium approximation of the black hole itself, we ignore the special relativistic Doppler effects due to the shrinkage of the surface of the black hole. 

\begin{figure}[t]
 \begin{center}
 \includegraphics[height=40mm]{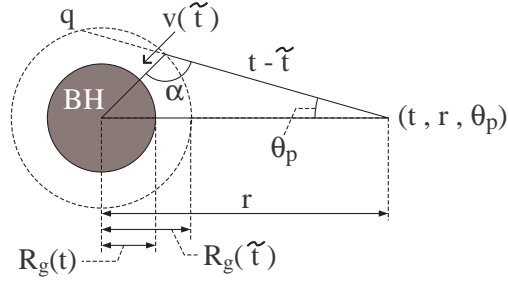}
 \end{center}
\caption{Retarded effect on the Hawking radiation.}
\label{pic-2}
\end{figure}

Hereafter, $r$ denotes the areal radius from the centre of the black hole. In order to take the retarded effect into account, consider a massless particle which has been emitted at the surface of the black hole at time $\rt$ and reaches the spatial point of $r$ at time $t \, ( > \rt )$. The emission time $\rt$ depends not only on the coordinates $(t,r)$, but also on the angle $\angl$ between the radial direction of $r$ and the momentum of the particle (see figure \ref{pic-2}). This emission time, $\rt(t,r,\angl)$, can be obtained as a root of the following equation,
\sikib
 R_g(\, \rt \,)^2
  = \left(\, t - \rt \, \right)^2 + r^2 - 2 \left(\, t - \rt \, \right) r\, \cos\angl \, .
\label{eq-entropy.emission}
\sikie
Note that this is the equation of degree six about $\rt$, and the appropriate root as the emission time $\rt(t,r,\angl)$ is the maximum root in the range, $0 \le \rt \le t$. Although another root may exist in this range, however such a root corresponds to the particle emitted at point $q$ in figure \ref{pic-2} which is obviously unphysical. And the other four roots of this equation may be complex valued ones. Further it is also obvious that the angle $\angl$ has an upper bound, $0 \le \angl \le \anglm(t,r)$. This upper bound is given by appendix \ref{app-upper} as (see equations \eqref{eq-upper.bound.1}, \eqref{eq-upper.bound.2} and \eqref{eq-upper.bound.3} )
\sikibnon
 \cos\anglm(t,r) &=& \frac{t - t_m(t,r)}{r} \qquad , \quad \text{for $r \le R_0$} \\
 \cos\anglm(t,r) &=& 
 \begin{cases}
    1 &, \quad \text{for $r > R_0$ and $t < r - R_0$} \\
  \dfrac{t^2 + r^2 - R_0^2}{2\, t\, r}
    &,\quad \text{for $r > R_0$ and $r - R_0 \le t \le \sqrt{r^2 - R_0^2}$} \\
  \dfrac{1}{r}\left[\, t - t_m(t,r) \,\right]
    &,\quad \text{for $r > R_0$ and $\sqrt{r^2 - R_0^2} < t$}
 \end{cases}
\sikienon
where $t_m(t,r)$ is the real valued root in the range, $0 \le t_m \le t$, of the equation,
\sikibnon
 r^2 = R_g(\, t_m \,)^2 + \left(\, t - t_m \,\right)^2 \, .
\sikienon
Then the distribution function of the free Hawking fields is given as
\sikib
 f(t , r ; \omega , \angl) =
  \begin{cases}
   \dfrac{1}{\exp\left[\, \omega / \tilde{T} \,\right] \pm 1}
                            &, \quad \text{for $\angl \le \anglm(t,r)$} \\
   0                        &, \quad \text{for $\angl > \anglm(t,r)$}
  \end{cases}
\label{eq-entropy.distribution}
\sikie
where $\tilde{T} = T_g(\, \tilde{t}(t,r,\angl) \,)$, $\omega = \left| \vec{p} \right|$, and the signatures ``$-$'' and ``$+$'' are respectively for bosons and fermions. Here note that, while the $t$- and $r$-dependence of $f$ expresses the spacetime dependence, the $\omega$- and $\angl$-dependence denotes the dependence on the momentum $\vec{p}$ of particles of the free Hawking fields. 

The entropy, $S_{nsi}(t)$, of the free Hawking fields can be obtained with substituting the distribution function \eqref{eq-entropy.distribution} into the place of $f_{\vec{p},\vec{x}}$ in formulae \eqref{eq-sst.ll.boson} and \eqref{eq-sst.ll.fermion},
\sikib
 S_{nsi}(t) = \int_{R_g(t)}^{t + R_0} dr\, 4 \, \pi\, r^2\, s_{nsi}(t,r) \quad , \quad
 s_{nsi}(t,r) = \frac{N}{2880 \, \pi} \int_{y_m(t,r)}^1 dy_p \,
                \frac{1}{R_g(\, \rt(t,r,y_p) \,)^3} \, ,
\label{eq-entropy.hr} 
\sikie
where the same integrals used in obtaining equations \eqref{eq-sst.entropy.rad} and \eqref{eq-sst.entropy.fermion} are also used, and we set $y_p = \cos\angl$, $y_m = \cos\anglm$ and $g_{\vec{p} , \vec{x}} = n_b$ and $n_f$ for bosons and fermions respectively. Here note that, since we assume that the emission of the free Hawking fields starts at $t=0$, the free Hawking fields fill the space in the range, $R_g(t) < r < t + R_0$ .

It is useful to consider here the validity of the nonequilibrium entropy \eqref{eq-entropy.hr}. We pay attention to the quasi-equilibrium approximation of the black hole itself. This approximation require, not only that the relation $t_{ev}/t_{bh} \gg O(1)$ holds, but also that the shrinkage speed of the surface of the black hole is less than unity, 
\sikib
 \left| \frac{d R_g(t)}{dt} \right| = \frac{R_0^3}{3\, t_{ev}} \frac{1}{R_g(t)^2} < 1 \quad \Rightarrow \quad t < t_{ev} - \frac{\sqrt{N}}{48 \sqrt{15 \, \pi}} \, .
\label{eq-entropy.range.slow}
\sikie
This denotes that, when $N < 48^2 \times 15 \pi \simeq 108573$, our entropy \eqref{eq-entropy.hr} is valid at least until one Planck time before the evaporation time $t_{ev}$.

\subsection{Total entropy}

From the above discussions, the total entropy of the NHR model which consists of the black hole and the free Hawking fields is expressed as $S_{tot}^{(min)}(t) = S_g(t) + S_{nsi}(t)$, where $S_g(t)$ is given by equations \eqref{eq-strategy.eos} and \eqref{eq-entropy.radius}, and $S_{nsi}(t)$ by equation \eqref{eq-entropy.hr}. In order to show the numerical plot of this total entropy, we normalise as follows,
\sikibnon
 \tau = \frac{t}{t_{ev}} \,\, , \,\, \rtau = \frac{\rt}{t_{ev}} \,\, , \,\,
 x = \frac{r}{R_0} \,\, , \,\,
 X_g(\tau) = \frac{R_g(t)}{R_0} = \left(\, 1 - \tau \,\right)^{1/3} \,\, , \,\,
 \Sigma_{tot}^{(min)}(\tau) = \frac{S_{tot}^{(min)}(t)}{S_{tot}^{(min)}(0)} \,\, , \,\,
 \sigma_{nsi}(\tau,x) = \frac{s_{nsi}(t,r)}{S_{tot}^{(min)}(0)} \,\, .
\sikienon
Then the normalised total entropy is
\sikib
 \Sigma_{tot}^{(min)}(\tau)
 =  X_g(\tau)^2
  + \frac{16}{9\, \lambda} \int_{X_g(\tau)}^{\lambda\,\tau + 1} dx 
    \int_{y_m(\tau,x)}^1 dy_p \, \frac{x^2}{X_g(\, \rtau(\tau,x,y_p) \,)^3} \, ,
\label{eq-entropy.total}
\sikie
where $\lambda$ is given by
\sikibnon
 \lambda = 1280 \, \pi \, \frac{R_0^2}{N} = \frac{t_{ev}}{R_0} = \frac{t_{ev}}{t_{bh}} \, .
\sikienon
Here note that, because the entropy $\Sigma_{tot}^{(min)}$ does not explicitly depend on $R_0$ (or $N$), then, once the value of $\lambda$ is fixed, the choice of $R_0$ (or $N$) is arbitrary with adjusting the value of $N$ (or $R_0$) appropriately to match with $\lambda$. And we assume $\lambda \gg O(1)$ as mentioned in section \ref{sec-entropy.model}. 

\begin{figure}[t]
\centerline{\includegraphics[height=70mm]{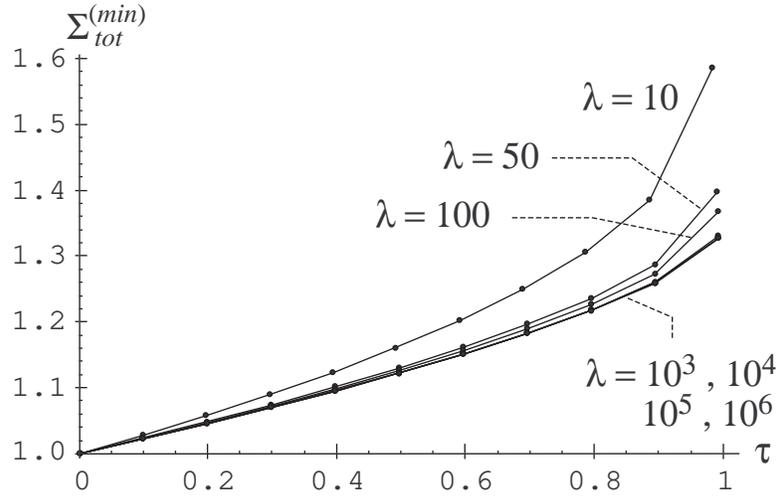}}
\caption{Time evolution of the normalised total entropy $\Sigma_{tot}^{(min)}(\tau) = S_{tot}^{(min)}(t)/S_{tot}^{(min)}$, where $\tau$ is the time normalised by the evaporation time $t_{ev}$. The plotted curves are converging as $\lambda \, (= t_{ev}/R_0)$ increases.}
\label{pic-total}
\end{figure}

\begin{figure}[t]
\centerline{\includegraphics[height=70mm]{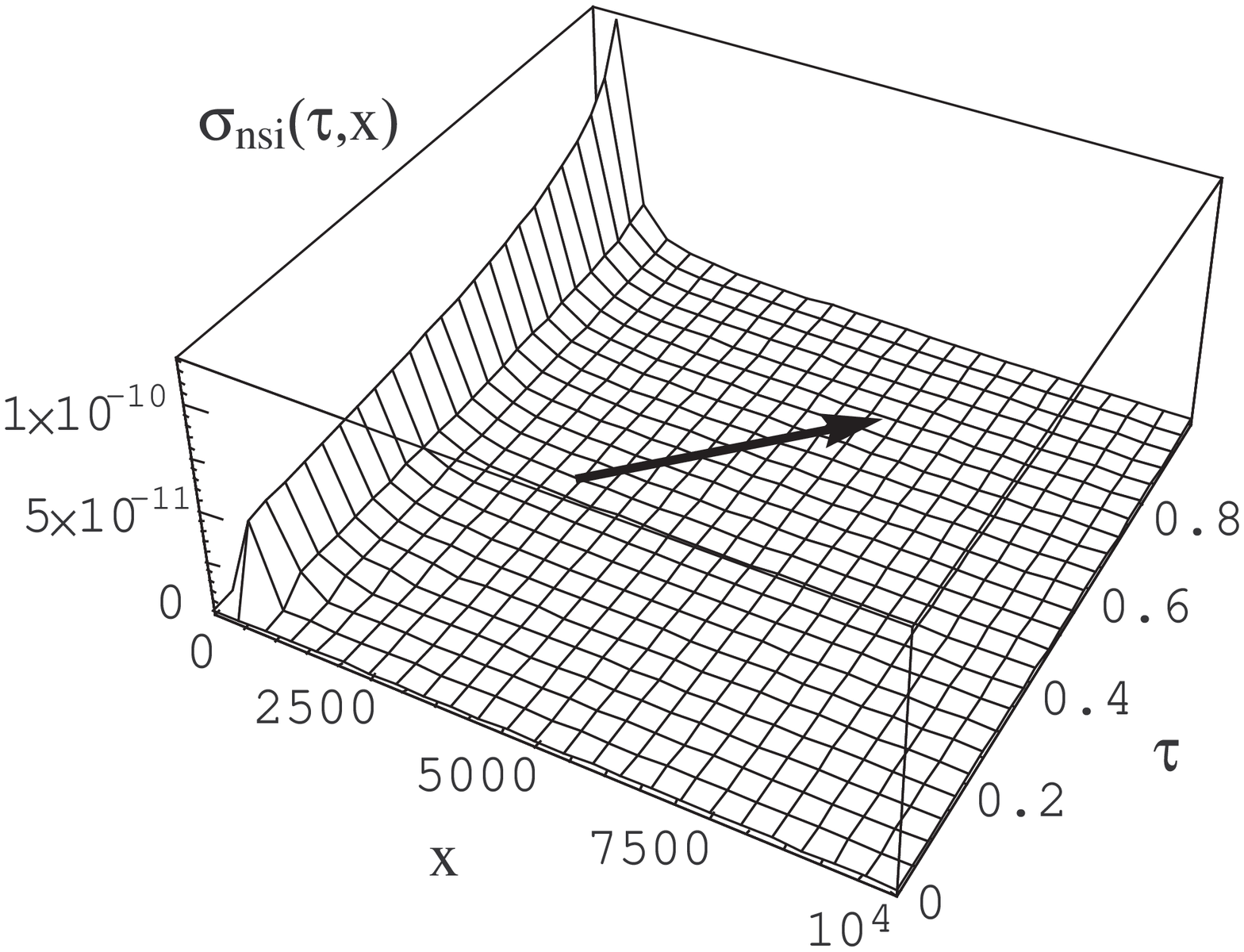} 
            \includegraphics[height=70mm]{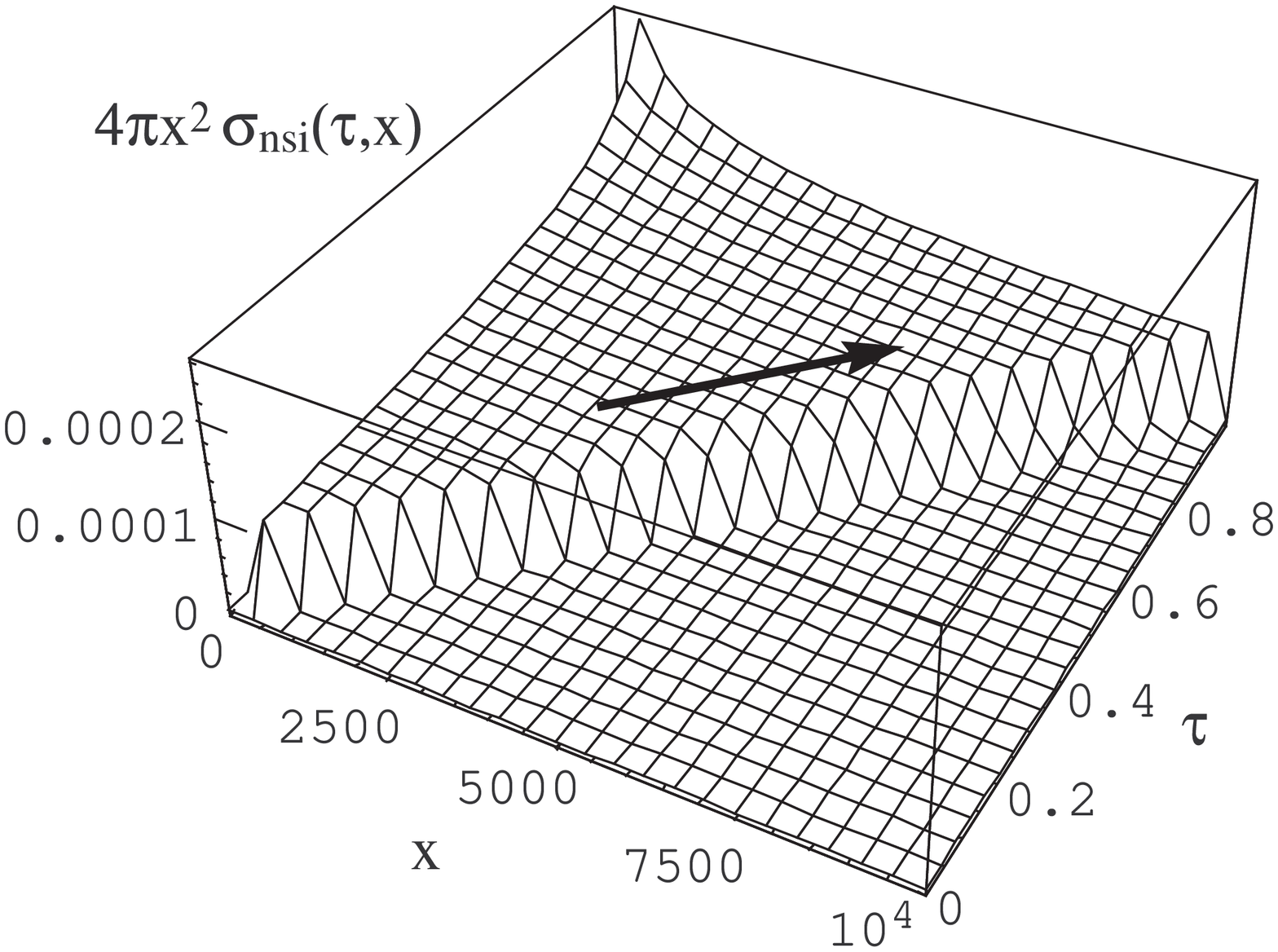}}
\caption{Spatial and temporal evolution of the entropy density $\sigma_{nsi}(\tau,x)$ and its spherical shell density $4 \pi x^2 \sigma_{nsi}(\tau,x)$ for $(\lambda = )\, t_{ev}/R_0 = 10^4$, where $\sigma_{nsi}(\tau,x) = s_{nsi}(t,r)/S_{tot}^{(min)}(0)$ is the normalised spatial density of the steady state entropy, $x = r/R_0$ is the normalised spatial distance and $\tau = t/t_{ev}$ is the normalised time. The arrows denote the direction of null world lines (geodesics).}
\label{pic-dens}
\end{figure}

The numerical results are shown in figures \ref{pic-total} and \ref{pic-dens}. The parameters $R_0$ and $N$ are chosen so that the relations $\lambda \gg O(1)$ and $N < 48^2 \times 15 \pi$ are satisfied, and that the time interval $0 \le \tau \le 0.999$ is included in the range \eqref{eq-entropy.range.slow}. The numerical plots are carried out in the interval $0 \le \tau \le 0.999$ with the {\it Mathematica} version 5.2.

Figure \ref{pic-total} shows the time evolution of the total entropy $\Sigma_{tot}^{(min)}$ for $\lambda = 10, 50, 100, 10^3, 10^4, 10^5$ and $10^6$
\footnote{Since the calculation has taken a very long time for double integral with {\it Mathematica}, figure \ref{pic-total} plots the values of $\Sigma_{tot}^{(min)}$ for $\tau = 0, 0.1, 0.2, \cdots , 0.9,$ and $0.999$ for each $\lambda$. Here note that any singular behaviour (oscillation, divergence and so on) of $\Sigma_{tot}^{(min)}$ is unexpected. Therefore, for the purpose of finding the monotone increasing nature of $\Sigma_{tot}^{(min)}$, it is enough to show the plot of some representative points like figure \ref{pic-total}.}. 
The plotted curves are converging as $\lambda$ increases. The plots for $\lambda \ge 10^3$ are almost coincident. What we can find with this figure is as follows.
\begin{itemize}
\item 
Because the plotted curves are converging, it is appropriate to conclude that the total entropy $\Sigma_{tot}^{(min)}$ is monotone increasing for $\lambda \gg O(1)$, and the GSL is well supported.
\item 
Since the normalised black hole entropy $X_g(\tau)^2$ is obviously monotone decreasing, the entropy of the free Hawking fields increases monotonically faster than the decrease of the black hole entropy.
\item 
It is suggested rather well that $\Sigma_{tot}^{(min)}(\tau)$ gives almost the same curves in $\Sigma$-$\tau$ graph for all values of sufficiently large $\lambda \gtrsim 10^3$, and the final value is approximately given by $\Sigma_{tot}^{(min)}(0.999) \simeq 1.33$.
\end{itemize}

The left panel in figure \ref{pic-dens} shows the normalised spatial density of the entropy of the free Hawking fields, $\sigma_{nsi}(\tau,x)$, and the right panel shows its normalised spherical shell density, $4 \pi x^2 \sigma_{nsi}(\tau,x)$. Figure \ref{pic-dens} is made with $\lambda = 10^4$, however the same behaviour is obtained for all values of sufficiently large $\lambda \gtrsim 10^3$. The arrows in figure \ref{pic-dens} denote the direction of the null world lines of massless particles. What we can find with this figure is as follows.
\begin{itemize}
\item 
While the entropy of the free Hawking fields spreads out to decrease the spatial density (left panel), the spherical shell density on each spherical shell remains constant while it spreads out into the empty space (right panel). This is consistent with the absence of the self-relaxation in the free Hawking fields. 
\item 
The entropy of the free Hawking fields at the surface of the black hole, which equals the amount of matter entropy emitted by the black hole at each moment, increases as the evaporation process proceeds. Here note that the the black hole temperature increases obviously. That is, the understanding of the GSL suggested in section \ref{sec-intro} can be extracted from figures \ref{pic-total} and \ref{pic-dens}.
\end{itemize}

\section{Summary and discussions}
\label{sec-sd}

When a black hole evaporates due to the Hawking radiation, the thermodynamic state of the Hawking field is in a nonequilibrium state. Then, referring to a recent development on the steady state thermodynamics for a radiation field, we have analysed explicitly the time evolution of a black hole evaporation process within the framework of the NHR model, and obtained the conclusion that the total entropy $S_{tot}^{(min)}$ increases monotonically for $( \lambda = )\, t_{ev}/R_0 \gg O(1)$, where $t_{ev}$ is the evaporation time and $R_0$ is the initial black hole radius. Therefore, as discussed in section \ref{sec-strategy}, we can obtain the understanding of the GSL in the context of the steady state thermodynamics as follows. The increase of the black hole temperature due to its negative heat capacity gives the increase of the entropy of the Hawking fields, and the rate of entropy increase of the Hawking fields is faster than the rate of entropy decrease of the black hole. That is, the self-interaction of the Hawking fields is not necessary for the validity of the GSL. 

Here we pay attention to the last stage of the evaporation process in the NHR model; the total entropy $S_{tot}$ increases in an accelerating fashion as shown in figure \ref{pic-total}. According to the relaxation process in the MS model which reaches a total equilibrium state as mentioned in section \ref{sec-sst} or shown in report \cite{ref-sst} in detail, one may think the accelerated increase of $S_{tot}^{(min)}$ is strange, because the black hole evaporation process in the framework of the NHR model is also described as the relaxation process of the whole system. However the accelerated increase of $S_{tot}^{(min)}$ never contradict the discussions in section \ref{sec-sst} and in report \cite{ref-sst}, because the black hole has the negative heat capacity as shown by equation \eqref{eq-strategy.capa}. The negative heat capacity is a peculiar property to the self-gravitating systems \cite{ref-sgs}. While the positive heat capacity of an ordinary system makes the entropy emission rate decrease to zero as the relaxation process proceeds, however the negative heat capacity of the black hole makes the entropy emission rate increase as the black hole evaporation process proceeds as shown by the right panel in figure \ref{pic-dens}. This increase of the entropy emission rate is the essence of our understanding of the GSL.

Here it is recalled that the negative heat capacity of a self-gravitating gas system (not of a black hole) causes the accelerated increase of its temperature, which is called the {\it gravothermal catastrophe} \cite{ref-sgs}. The gravothermal catastrophe is the phenomenon suggested by the ordinary equilibrium thermodynamics and statistical mechanics. Hence the accelerated increase of the total entropy of the black hole and the Hawking fields shown in figure \ref{pic-total} may be interpreted as a nonequilibrium version of the gravothermal catastrophe.

In this paper, the monotone increasing nature of the total entropy $S_{tot}^{(min)}$ is concluded by the numerical results shown in figure \ref{pic-total}. Therefore, correctly speaking, our conclusion is not analytically completely supported at present. If the inequality $dS_{tot}^{(min)}/dt > 0$ will be obtained analytically, our understanding of the GSL will have more rigid mathematical reasoning. Further our discussion given in this paper may be understood as a new proof of the GSL, $dS_{tot} > 0$, according to inequality \eqref{eq-strategy.ineq}. 

We turn our discussion to the validity of the assumption $\lambda = t_{ev}/t_{bh} \gg O(1)$, which guarantees the validity of the quasi-equilibrium approximation of the black hole. In semi-classical treatment, the black hole temperature is less than the Planck temperature. Then the exotic particles are not expected to be created in the Hawking radiation, and only the ordinary well-known standard particles will be created. Hence we can expect that the number of species of the Hawking fields ($\sim N$) is less than $108573 = O(10^5)$ (see inequality \eqref{eq-entropy.range.slow} ). For the case $N \sim O(10^5)$, the assumption $\lambda \gg O(1)$ is valid for a black hole of initial radius $R_0 \gtrsim O(10)$ in Planck unit. And if we estimate the order of $N$ by the ordinary standard particles (three generations of quarks and leptons and their inner states, gauge particles of four fundamental interactions and their helicities), then we find $N \sim O(10)$ and the assumption $\lambda \gg O(1)$ is valid for a black hole of initial radius $R_0 \gtrsim O(10^{-1})$ in Planck unit. Hence, when no exotic particle appears in the Hawking radiation, our discussion based on the NHR model seems to be valid for a semi-classical black hole of initial radius $R_0 > 1$ until one Planck time before the evaporation time $t_{ev}$.

Here we notice that, for the case $N \sim O(10)$, the order relation $\lambda \gtrsim O(10^3)$ corresponds to the black hole of initial radius $R_0 \gtrsim O(10)$. Therefore we find from figure \ref{pic-total} that, for the black hole of $R_0 \gtrsim O(10) \, (\Rightarrow \lambda \gtrsim O(10^3) )$, the normalised total entropy $\Sigma_{tot}^{(min)}(\tau)$ evolves in almost the same increasing fashion and reaches almost the same final value $\simeq 1.33$. Further it should be noted here that, even if we consider the very last stage of the evaporation process where the quasi-equilibrium approximation of the black hole breaks down and the evaporation may proceed to a quantum gravitational process, it is reasonable to expect that the total entropy after the quantum evaporation process finished should be greater than the total entropy before the evaporation process proceeds to the quantum one, since the well-defined entropy has to be a non-decreasing quantity. Hence, because inequality \eqref{eq-strategy.ineq} indicates that $S_{tot}^{(min)}$ is the lower bound of the total entropy of the whole system, it is suggested that, when the black hole evaporates in the empty space with emitting any self-interacting Hawking field, the final value of the total entropy $S_{tot}$ (not the lower bound $S_{tot}^{(min)}$) should be larger than $1.33 \times S_g(0)$. 

Finally we discuss about the strategy to reach the understanding of the GSL in the context of the steady state thermodynamics. The strategy explained in section \ref{sec-strategy} has two key assumptions remained to be proven, the additivity of the entropy \eqref{eq-strategy.total.general} and the existence of a well-defined nonequilibrium entropy $S_m$ of an arbitrary self-interacting Hawking field. However, these two key assumptions seem to be reasonable as are discussed in the following two paragraphs. 

On the validity of the additivity \eqref{eq-strategy.total.general}, we refer the present status of the study on the laboratory system which has self-interactions but not self- and external-gravitational interactions. For example, consider that the system is in a nonequilibrium state which is not so far from an equilibrium state that the heat flux in that system is not extremely strong. Then the so-called extended irreversible thermodynamics \cite{ref-eit} gives a well-defined nonequilibrium entropy up to the second order in the expansion by the heat flux. Further the nonequilibrium entropy satisfies the additivity as in the ordinary equilibrium thermodynamics. Therefore, we can expect that, when the equilibrium entropy satisfies the additivity even with the gravitational effects, the nonequilibrium entropy also satisfies the additivity. Hence, due to the additivity of the equilibrium entropies of the black hole and the Hawking field \cite{ref-add}, it seems reasonable to assume the additivity \eqref{eq-strategy.total.general}. 

Next we consider the validity of the existence of a well-defined nonequilibrium entropy $S_m$ of an arbitrary self-interacting Hawking field. As mentioned in the previous paragraph, the nonequilibrium entropy has already been defined well up to the second order in the expansion by the heat flux \cite{ref-eit}. Further, on the other hand, the evaporation time $t_{ev}$ of a black hole has an extremely long time scale, $\lambda = t_{ev}/t_{bh} \gg O(1)$, as discussed above. Therefore it seems appropriate that the black hole evaporation proceeds so slowly that the nonequilibrium state of the self-interacting Hawking field is not so far from an equilibrium state that the expansion of the thermodynamic quantities by the heat flux up to the second order can be a good approximation. Hence it is reasonable to assume the existence of a well-defined nonequilibrium entropy of an arbitrary self-interacting Hawking field.

\appendix
\section{Black hole evaporation and star formation process}
\label{app-star}

This appendix discusses in a simple way about the similarity between the black hole evaporation process and the star formation process. When an interstellar gas collapses to form a star, it is commonly believed that there arises the increase of the net entropy of the system which consists of the collapsing gas and the radiated matters from the gas. However the {\it self}-gravitational effect of the collapsing gas causes the decrease of the entropy of the {\it collapsing} gas itself. It is briefly explained as follows.

Since the pressure of the gas at its surface is zero, the loss of energy $\Delta E$ (par a unit time) of the collapsing gas by the radiation of energy is actually the loss of heat due to the first law of the thermodynamics,
\sikib
 \Delta E \sim T \, \Delta S \, 
\label{eq-star.1}
\sikie
where $T$ is the temperature of the collapsing gas, and $\Delta S$ is the {\it loss} of entropy (par a unit time). The energy loss $\Delta E$ can be interpreted as the minus of the luminosity $L$ of the collapsing gas, $L = - \Delta E$. Then, with the assumption of the local mechanical and thermal equilibrium of the collapsing gas at each moment of its collapse, relation \eqref{eq-star.1} can be rewritten more exactly to the following form \cite{ref-sgs},
\sikibnon
 L = - \frac{d [ \, U + \Omega \,]}{dt} \sim - T \Delta S \, ,
\sikienon
where $U$ is the total internal energy of the collapsing gas, and $\Omega$ is the total self-gravitational potential given by
\sikibnon
 \Omega = \int_0^M \, dm \,\left( - \frac{G\, m}{r(m)} \, \right) \, ,
\sikienon
where $G$ is Newton's constant, $M$ is the mass of the collapsing gas, and the radial distance from the centre of the gas $r(m)$ is expressed as the function of mass $m$ inside the sphere of radius $r$. 

From above, it is recognised that the radius $r(m)$ becomes smaller as the gas collapses, then the self-gravitational potential $\Omega$ decreases to result in the radiation of energy $L >0$ and the loss of the entropy $\Delta S < 0$. Therefore, if we consider not a collapsing but an expanding self-gravitating gas, the radius $r(m)$ increases to result in the increase of entropy $\Delta S >0$. Here note that the Hawking field in the black hole evaporation process corresponds not to the collapsing gas but to the the expanding matters radiated by the collapsing gas. Hence it seems reasonable to assume that the entropy of a self-gravitating Hawking field is larger than that of a non-self-gravitating Hawking field.

\section{Upper bound of $\angl$}
\label{app-upper}

\begin{figure}[t]
 \begin{center}
 \includegraphics[height=40mm]{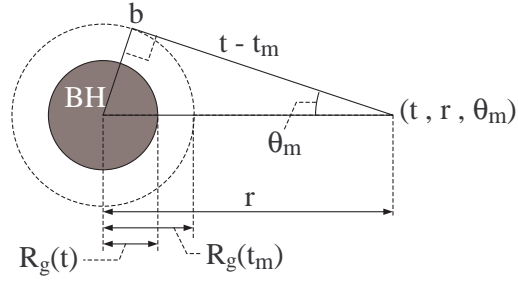}
 \end{center}
\caption{Upper bound of $\angl$ for $r < R_0$, or $r > R_0$ and $t > \sqrt{r^2 - R_0^2}$.}
\label{pic-3}
\end{figure}

This appendix derives the upper bound $\anglm$ of the angle $\angl$ which appears in equation \eqref{eq-entropy.emission}. In order to find the explicit expression of $\anglm$, look at the photons emitted by the black hole at the initial time $t=0$, which we call the initial photons. Obviously, we have $\anglm = 0$ for $t < r - R_0$, since no photons has been reached the point. The initial photons emitted in radial direction form the boundary of the region which the free Hawking fields fill. The initial photons emitted in off-radial directions propagate behind the initial photons emitted in radial directions. Therefore, the initial photons reach a spatial point of radial distance $r  > R_0$ in the time interval, $r - R_0 \le t \le \sqrt{r^2 - R_0^2}$. Within this time interval, the upper bound $\anglm$ is given by the initial photon, which is obtained with setting $\rt = 0$ in figure \ref{pic-2}. Then we find
\sikib
 \cos\anglm(t,r)
 = \frac{t^2 + r^2 - R_0^2}{2\, t\, r} \quad , \quad
   \text{for $r > R_0$ and $r - R_0 \le t \le \sqrt{r^2 - R_0^2}$} \, .
\label{eq-upper.bound.1}
\sikie
Further for $t > \sqrt{r^2 - R_0^2}$, the upper bound $\anglm$ is not given by any initial photon but by the photon emitted at the point $b$ at the time $t_m$ shown in figure \ref{pic-3}. Then we find
\sikib
 \cos\anglm(t,r)
 = \frac{t - t_m(t,r)}{r} \quad , \quad
   \text{for $r > R_0$ and $\sqrt{r^2 - R_0^2} < t$} \, ,
\label{eq-upper.bound.2}
\sikie
where the time $t_m(t,r)$ is a real valued root of the equation, $r^2 = R_g(\, t_m \,)^2 + \left(\, t - t_m \,\right)^2$. This is the equation of degree six about $t_m$. The appropriate root for $t_m$ should be in the range, $0 \le t_m \le t$. This root may be degenerated, since the trajectory of this photon is tangent to the sphere of radius $R_g( t_m )$, and the other four roots may be complex valued ones. Finally turn to the spatial point of radial distance $r \le R_0$. It is obvious for this spatial point that the upper bound $\anglm$ is also given by formula \eqref{eq-upper.bound.2},
\sikib
 \cos\anglm(t,r)
 = \frac{t - t_m(t,r)}{r} \quad , \quad
   \text{for $r \le R_0$} \, .
\label{eq-upper.bound.3}
\sikie

\end{document}